\documentclass[aps,reprint,amsmath,amssymb, superscriptaddress]{revtex4-1}
\pdfoutput=1
\usepackage[T1]{fontenc}
\usepackage[latin9]{inputenc}
\usepackage[pdftex]{graphicx}
\usepackage{color, soul}

\begin{document}

\renewcommand*\rmdefault{bch}\normalfont\upshape
\rmfamily
\section*{}
\vspace{-1cm}

\title{Using Depletion to Control Colloidal Crystal Assemblies of Hard Cuboctahedra}
\author{Andrew S. Karas}
\author{Jens Glaser}
\affiliation{Department of Chemical Engineering, University of Michigan, Ann Arbor}
\author{Sharon C. Glotzer}
\email{sglotzer@umich.edu}
\affiliation{Department of Chemical Engineering, University of Michigan, Ann Arbor}
\affiliation{Department of Materials Science and Engineering, University of Michigan, Ann Arbor}
\date {\today}

\begin{abstract}
\vspace{.3cm}

Depletion interactions arise from entropic forces, and their ability to induce aggregation and even ordering of colloidal particles through self-assembly is well established, especially for spherical colloids. We vary the size and concentration of penetrable hard sphere depletants in a system of cuboctahedra, and we show how depletion changes the preferential facet alignment of the colloids and thereby selects different crystal structures. Moreover, we explain the cuboctahedra phase behavior using perturbative free energy calculations. We find that cuboctahedra can form a stable simple cubic phase, and, remarkably, that the stability of this phase can only be rationalized by considering the effects of both the colloid and depletant entropy. We corroborate our results by analyzing how the depletant concentration and size affect the emergent directional entropic forces and hence the effective particle shape. We propose the use of depletants as a means of easily changing the effective shape of self-assembling anisotropic colloids.
\vspace{.7cm}

\end{abstract}

\maketitle
\nopagebreak









\section{Introduction}

New, programmable materials can be realized from nanoparticles that form ordered solids to maximize their entropy. Recent studies have investigated how changes to particle shape affect the solids formed via self-assembly \cite{Damasceno2012b, vanAnders2014e, Damasceno2012a, Gantapara2013, Ni2012a}. In such systems, directional entropic forces emerge upon crowding so that particles adopt locally dense configurations to maximize the system entropy \cite{vanAnders2014d}. In binary systems, the depletion interaction, in which the osmotic pressure of small polymers induces an effective attraction between colloids, provides another route to induce order via entropy \cite{AsakuraOosawa}. When combined with shape, depletion is a powerful tool to control structure formation because it promotes facet-specific alignment of anisotropic particles even for low concentrations of colloids \cite{Mason2002}. For example, in systems of roughened platelets, adjusting the depletant size allows one to selectively suppress or enhance specific bonding \cite{Zhao2008}. In three-dimensional crystalline assemblies of metallic nanoparticles, depletants have been used to control lattice spacings \cite{GlotzerMirkin2013} and to stabilize exotic structures instead of densest packings \cite{Henzie2012b}. The swelling of thermo-sensitive depletants causes reversible transformations in cube-like superballs between a square lattice and a canted phase in a quasi-two dimensional system \cite{Rossi2015}. By calculating the probability of binding, or facet alignment, for hard facetted spheres with depletants, van Anders et al. \cite{vanAnders2014d} demonstrated how changing depletant concentration and facet size changes the strength of emergent directional entropic forces.

Cuboctahedra belong to a shape family in which a cube can be truncated into an octahedron. Simulations predict that cuboctahedra form a rotator body centered cubic (BCC) crystal at intermediate pressures while the cuboctahedra densest packing (CODP) formed at high pressures can be viewed as a sheared BCC crystal with a rhombohedral primitive cell \cite{Gantapara2013}. Both rotator and sheared BCC (sBCC) structures are observed in self-assemblies of perfect octahedra, though no previous studies have shown hard cuboctahedra forming a simple cubic (SC) phase, as is known to form from cubes. Experimentally, silver nanoparticles with cuboctahedral shape have been grown with diameters of 150-200 nm \cite{Tao2006}. Henzie et al.~\cite{Henzie2012b} used such particles in sedimentation-driven self-assembly studies and were able to form the densest-packed structure as well as a face-centered tetragonal (FCT) structure with depletion.

Here we show how depletion-induced effective shape change can control the self-assembly of polyhedral particles, using the specific example of cuboctahedra. Monte Carlo simulations that treat depletants as penetrable hard spheres demonstrate the ability to change the self-assembly of cuboctahedra into different colloidal crystals. We report that depletion can lead to the assembly of a simple cubic phase. Extending previous models for hard spheres \cite{Lekkerkerker1992}, we incorporate the contributions of both colloid and depletant entropy into a consistent thermodynamic picture of the phase behavior of anisotropic colloids with depletants. Remarkably, we find that, although the contribution to the free energy is dominated (as expected) by the depletant entropy, the {\it colloid} entropy contribution must be included to rationalize the stability of the SC phase. We further show how varying depletion size and concentration changes the directional entropic forces acting upon the cuboctahedra, and we argue that these changes to entropic forces are akin to effective shape changes with anisotropic particles.

\begin{figure}
\centering
\includegraphics[width=\columnwidth]{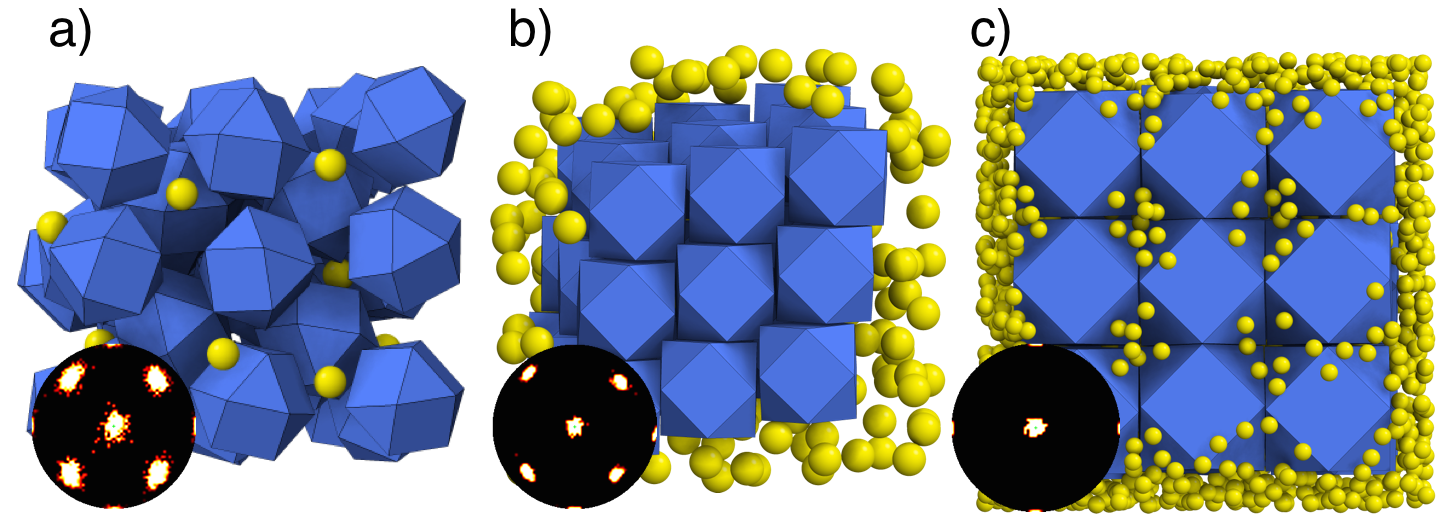}
\caption{Our proposed mechanisms that lead to the different crystal structures as depletant size and concentration is changed. Representative bond order diagrams are also shown. \textbf{(a)} In the rotator BCC phase, the depletant concentration is low and the depletants (yellow) sit in random void spaces throughout the structure. \textbf{(b)} In the sBCC phase, the cuboctahedra (blue) will pack densely and the depletants exist outside of the crystal, providing osmotic pressure which causes the dense packing. \textbf{(c)} In the simple cubic phase, sufficiently small depletants can sit in octahedral void spaces throughout the SC crystal in addition to sitting outside of the crystal in large pockets of free volume.}
\label{schematic}
\end{figure}

\section{Methods}

To study the behavior of cuboctahedra with depletants, we simulated their assembly across a range of depletant and colloid densities, for two depletant sizes.  We used a recently developed Monte Carlo (MC) simulation method that implicitly treats arbitrarily shaped hard particles in osmotic equilibrium with a reservoir of penetrable hard sphere depletants \cite{JensMethods}. This method is built on top of an existing parallel scheme for hard particle MC simulations \cite{HPMCarxiv, Anderson2008, Anderson2013a, HoomdWebsite}. In these simulations, we consider a semigrand ensemble where the system is kept at a constant volume with colloid density \(\phi_{\mathrm{col}}\) and depletant reservoir density \(\phi_{\mathrm{dep}}^{\mathrm{r}}\). The depletant size is set by a parameter \(q = \sigma_{\mathrm{dep}}/\sigma_{\mathrm{col}}\), the ratio of depletant radius to colloid circumsphere radius. Our MC method only stores information on the colloids; it accounts for depletion effects by sampling the free volume change in the local environment of a moved colloid in a statistically exact fashion.

We use a perturbative \textit{ansatz} known from Free Volume Theory (FVT) in order to predict coexistence of different cuboctahedra phases. Our implementation follows the framework of previous implementations \cite{Lekkerkerker1992, Dijkstra1999, Oversteegen2005}, although a key difference is that we rely on numerical calculations of the cuboctahedra free energy whereas previous implementations of FVT have focused on hard sphere systems where there are well developed analytical expressions for the equation of state and free volume.

FVT makes the approximation that the system's total free energy can be accurately approximated by the Helmholtz free energy of the depletant-free system plus a depletant perturbation term that relies only on the depletant number density $n_{\mathrm{dep}}^{\mathrm{r}} = \phi_{\mathrm{dep}}^{\mathrm{r}}/v_{\mathrm{0, dep}}$, the ratio of free volume to total system volume, $ \alpha = V_{\mathrm{free}}/V$, and the thermal energy scale, $kT$ (the Boltzmann constant multiplied by temperature). In other words,

\begin{equation}
F (N_{\mathrm{col}}, \mu_{\mathrm{dep}}, V, T) \approx A(N_{\mathrm{col}}, V, T) - n_{\mathrm{dep}}^{\mathrm{r}} \, \alpha V \, kT
\label{eq:fvt}
\end{equation}

We use two different means of computing the Helmholtz free energy of the unperturbed system. Equation of state data was used to determine the free energies of the fluid and rotator BCC phases, and Frenkel-Ladd integration \cite{Frenkel1984a, Haji-Akbari2011h} was employed to calculate free energies for the bulk SC and sBCC crystals. For the depletant perturbation term, we calculated the free volume fraction $ \alpha $ as a function of colloid phase behavior and concentration using MC integration. With expressions for all terms of the free energy, we used common tangent construction to find the densities of coexisting phases in the $ (\phi_{\mathrm{col}}, \, \phi_{\mathrm{dep}}^{\mathrm{r}}) $ plane. For further details on our calculation of phase behavior, see the Supplementary Information.

We also calculate the potential of mean force and torque (PMFT) in systems of cuboctahedra with depletants. These calculations were performed on dense fluid systems and they allow us to visualize the directional entropic forces (DEFs) that will ultimately result in solid formation \cite{Damasceno2012b, vanAnders2014e, vanAnders2014d, Harper2015}.

\section{Results and Discussion}

\begin{figure}[t!]
\centering
\includegraphics[width=\columnwidth]{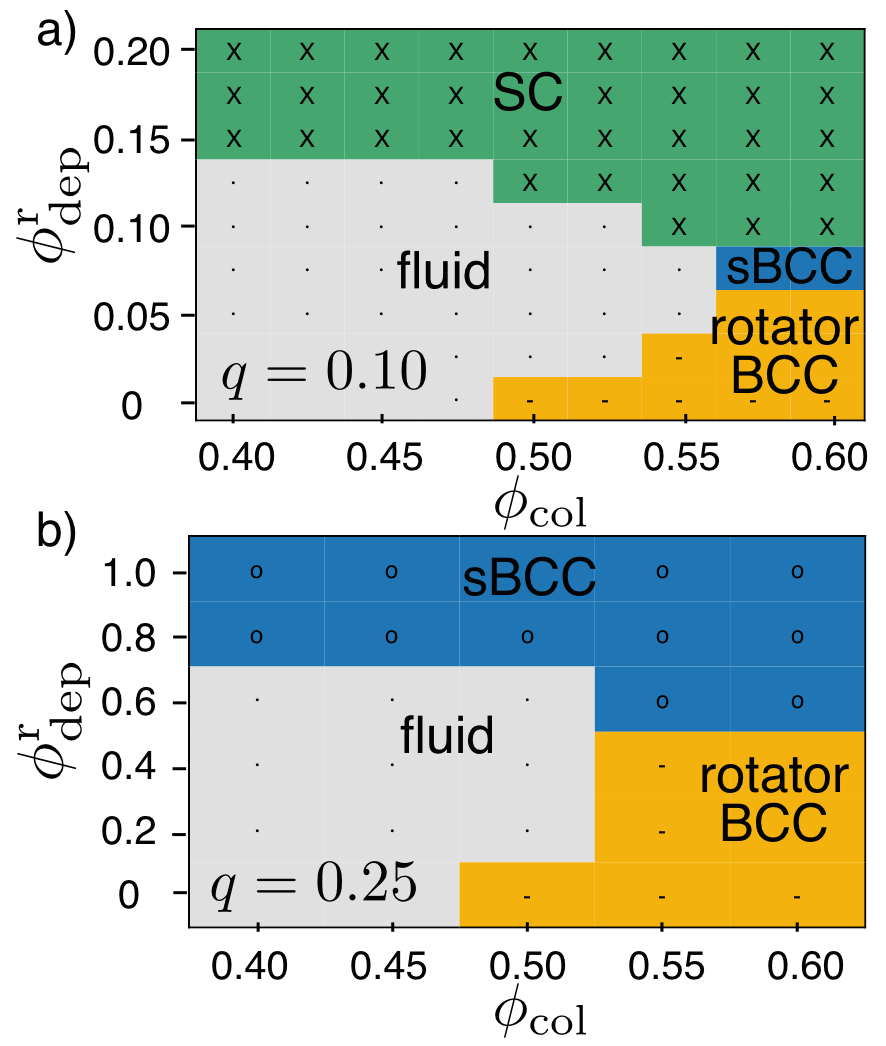}
\caption{ \textbf{(a \& b)} Observed phase behavior of cuboctahedra in self-assembly simulations with depletants at size ratios of \textbf{(a)} \(q=0.10\) and \textbf{(b)} \(q=0.25\) for various colloid densities (\(\phi_{\mathrm{col}}\)) and depletant concentrations (\(\phi_{\mathrm{dep}}^{\mathrm{r}}\)).}
\label{phase_diagram}
\end{figure}

Figures \ref{schematic} and \ref{phase_diagram} show examples of the ordered structures formed in our self-assembly simulations as we adjusted \(\phi_{\mathrm{col}}\), \(\phi_{\mathrm{dep}}^{\mathrm{r}}\), and \(q\), and the phase diagram, respectively. We observe three types of ordered structures: a rotator BCC crystal, a sheared BCC crystal with similar features as the CODP, and an SC crystal. We identify these structures based on visual inspection and their bond order diagram (Fig.~\ref{schematic}, insets). When the sBCC or SC crystals formed, the system phase separated into a dense solid and a less dense fluid, while rotator BCC formation led to a single phase. For any point in Figure~\ref{phase_diagram} labeled as a fluid, no ordered phase formed after $2.5\times 10^7$ MC sweeps. In contrast, systems often formed solids after approximately $5\times 10^6$ MC sweeps. We first probed the phase behavior by running two independent simulations at each state point with system sizes of \(N_{\mathrm{col}}=512\). We then confirmed these results by running four replicates at state points along the phase boundaries with system sizes \(N_{\mathrm{col}}=1000\). With depletant size ratios \(q=0.10\) and \(0.25\), the type of crystal formed at high depletant concentrations is consistent among replicates. Overall, we observe the SC crystal to be the only ordered phase formed in simulations with \(q=0.10\) and \(\phi_{\mathrm{dep}}^{\mathrm{r}} \geq 0.10\), and the sBCC crystal to form with \(q=0.25\) and \(\phi_{\mathrm{dep}}^{\mathrm{r}} \geq 0.60\).

From the phase diagram, we identify three regimes that lead to the cuboctahedra ordering. These different behaviors are visually represented with explicit depletants in Figure \ref{schematic}. The cuboctahedra behave as they would in a one-component system regardless of the depletant size when \(\phi_{\mathrm{dep}}^{\mathrm{r}}\) is sufficiently small (Fig.~\ref{schematic}\textbf{a}). Under these conditions, there is no preferential facet alignment as the particles are able to freely rotate. When \(\phi_{\mathrm{dep}}^{\mathrm{r}}\) increases, the assembly behavior of the cuboctahedra depends on the depletant size in the following way. With large depletants \(q=0.25\), the cuboctahedra behave as they would in the high pressure limit for one-component systems and pack into a dense sBCC crystal (Fig.~\ref{schematic}\textbf{b}). This dense arrangement contains significant contact among facets wherein both the square and triangular facets of a cuboctahedra are in contact with neighboring particles. With small depletants \(q = 0.1\), however, the cuboctahedra form an SC lattice that maximizes facet-to-facet alignment of the large square facets  (Fig.~\ref{schematic}\textbf{c}). This structure contains octahedral void space so that there is significant free volume available to the depletants within the crystal lattice. We do not observe the FCT crystal reported by Henzie et al.~\cite{Henzie2012b} even when using depletants of the same size as in their experiments (\(q=0.07\)). For this depletant size, we instead observe SC crystal formation  \footnote{See Supplementary Information for simulations with $q=0.07$ depletants.} which leads us to suspect other forces such as gravity contribute to the stabilization of cuboctahedra into an FCT crystal.

We validate our phase diagrams from simulations in Figure~\ref{phase_diagram} by predicting phase behavior using the perturbation relationship in Equation 1. With this expression, which accounts for contributions from both the colloid and depletant entropy, we are able to predict whether the system can minimize its free energy by remaining in one single phase or separating into two separate phases that are at mechanical and chemical equilibria. We include example free energy plots at specific depletant concentrations in the Supplementary Information.

\begin{figure}[t!]
\centering
\includegraphics[width=\columnwidth]{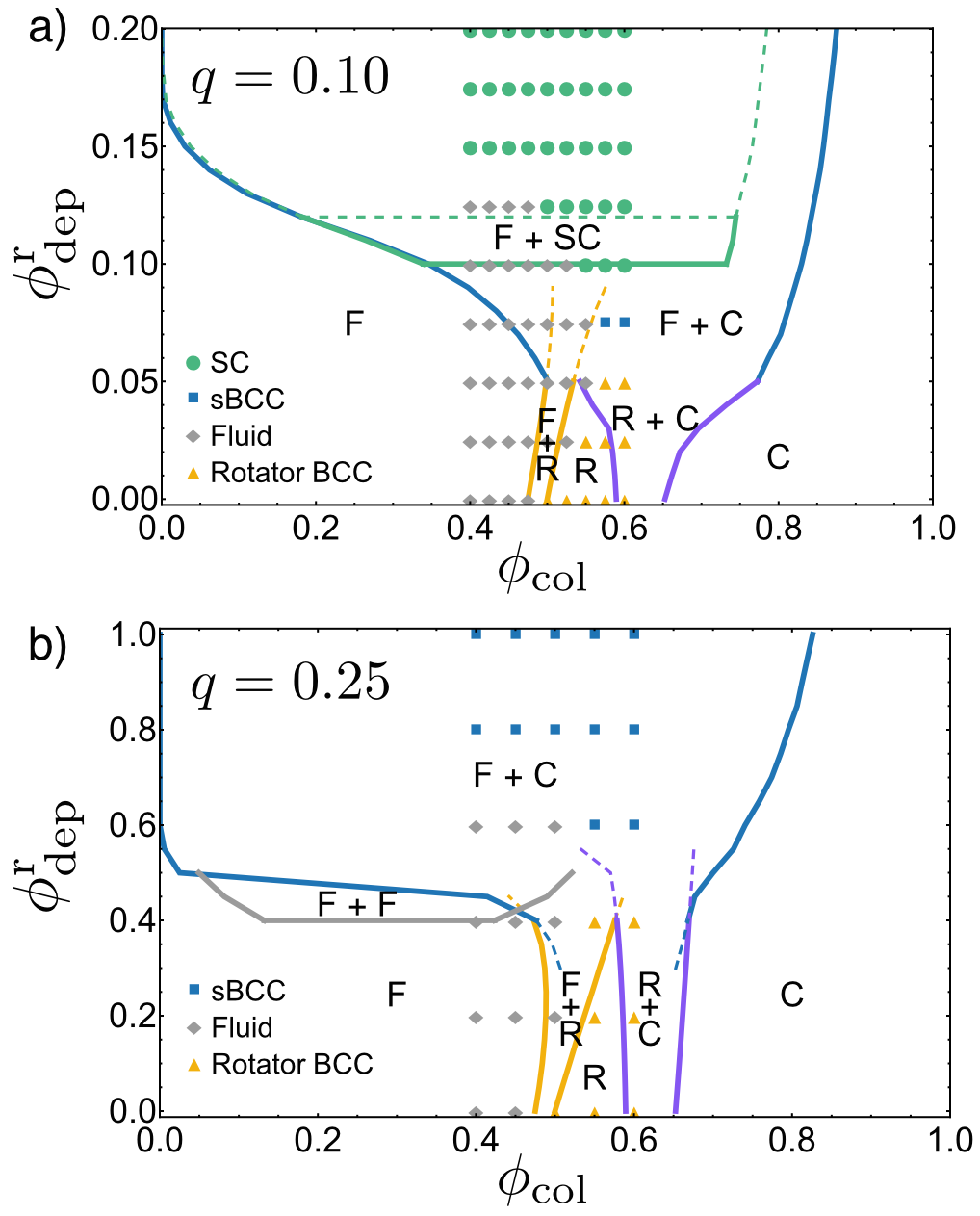}
\caption{Phase diagrams of cuboctahedra with penetrable hard sphere depletants at size ratios \textbf{(a)} $q=0.10$  and \textbf{(b)} $q=0.25$ as a function of $\phi_{\mathrm{col}}$ and $\phi_{\mathrm{dep}}^{\mathrm{r}}$. Predictions of the phase behavior from free energy calculations are noted with via text and the curves. The observed phase behavior from self-assembly simulations are noted with markers. F denotes a fluid phase, R denotes a rotator BCC phase, C denotes the sBCC phase, and SC denotes the simple cubic phase. Regions of coexistence between two phases are noted with a +. Metastable phases are denoted with dashed lines. In \textbf{(a)}, our calculations predict the F+SC coexistence to be stable for $ 0.10 \leq  \phi_{\mathrm{dep}}^{\mathrm{r}} < 0.12 $ and metastable for $\phi_{\mathrm{dep}}^{\mathrm{r}} \geq 0.12 $.}
\label{coexistence_diagram}
\end{figure}

Figure \ref{coexistence_diagram} shows phase diagrams as a function of $\phi_{\mathrm{col}}$ and $\phi_{\mathrm{dep}}^{\mathrm{r}}$ for both depletant sizes of interest. We denote single phase regions using a single letter, and we show regions of coexistence as those contained within the solid curves between two regions. In the two-phase regions, the system is able to minimize its free energy by separating into the two phases of different densities. We additionally represent phases predicted to be metastable using dashed lines. The observed phase behavior represented in Figure \ref{phase_diagram} is also included as symbols.

In both Figures \ref{coexistence_diagram}\textbf{(a \& b)}, we find the same behavior at low depletant concentrations: as the colloid concentration increases, the system transitions from a fluid to a rotator BCC crystal, and then to the sBCC crystal. The behavior with $\phi_{\mathrm{dep}}^{\mathrm{r}} = 0$ is in agreement with prior calculations of cuboctahedra phase behavior in Ref. \citenum{Gantapara2013}. The fact that the F+R coexistence curve slants to the right with increasing $\phi_{\mathrm{dep}}^{\mathrm{r}}$indicates that a higher $\phi_{\mathrm{col}}$ is needed for nucleation of the rotator BCC phase, which is what we observe from simulation.

For $q=0.10$ (Fig.~\ref{coexistence_diagram}\textbf{a}), increasing the depletant concentration so that $ 0.05 \leq \phi_{\mathrm{dep}}^{\mathrm{r}} < 0.10$ will destabilize the rotator BCC crystal and bring about coexistence between a fluid and the sBCC crystal. When $ 0.10 \leq  \phi_{\mathrm{dep}}^{\mathrm{r}} < 0.12 $, we predict the fluid-SC coexistence to be stable as compared to the fluid-sBCC coexistence. Further increases to $ \phi_{\mathrm{dep}}^{\mathrm{r}} $ cause the fluid-SC coexistence to become metastable under our predictions. We note that we do not observe the predicted stable sBCC phase in simulations at $\phi_{\mathrm{dep}}^{\mathrm{r}} > 0.12$, which could be a consequence of metastability or because the perturbative estimate for the metastability limit only represents a lower bound.

The reason for this predicted fluid-SC metastability is that high depletant concentrations favor denser packings. Though the sBCC crystal lacks  $V_{\mathrm{free}}$ within the crystal, the fact that it can pack more densely than the SC crystal means that it can maximize the amount of $V_{\mathrm{free}}$ outside of the crystal phase (Fig. \ref{schematic}\textbf{b}). These results indicate that performing a ground state analysis, i.e. the formal limit $n_{\mathrm{dep}}^{\mathrm{r}}\to\infty$ that only incorporates the depletant entropy as based on the amount of $V_{\mathrm{free}}$ in the system, will not properly resolve the phase behavior observed in our simulations. It is only by considering the entropy of the colloids that we are able to predict the stability of the simple cubic crystal. This contrasts recent work on the quasi-2D assembly of rounded cubes wherein a ground state analysis was sufficient in predicting phase behavior \cite{Rossi2015}.

For $q=0.25$ (Fig.~\ref{coexistence_diagram}\textbf{b}), the rotator BCC crystal destabilizes when $ \phi_{\mathrm{dep}}^{\mathrm{r}} > 0.40 $, and the fluid-sBCC coexistence is the dominant phase for high depletant concentrations. Our simulations and calculations differ on the critical $\phi_{\mathrm{dep}}^{\mathrm{r}}$ above which the fluid-sBCC coexistence is observed. Previous comparisons of FVT to simulations show that FVT tends to underestimate such critical depletant concentrations \cite{Bolhuis2002}. Our calculations also predict a gas-liquid coexistence for a narrow range of depletant densities. We confirm this phase behavior with simulations in the Gibbs ensemble which we describe in the Supplementary Information.

\begin{figure}[t!] 
\centering
\includegraphics[width=1.0\columnwidth]{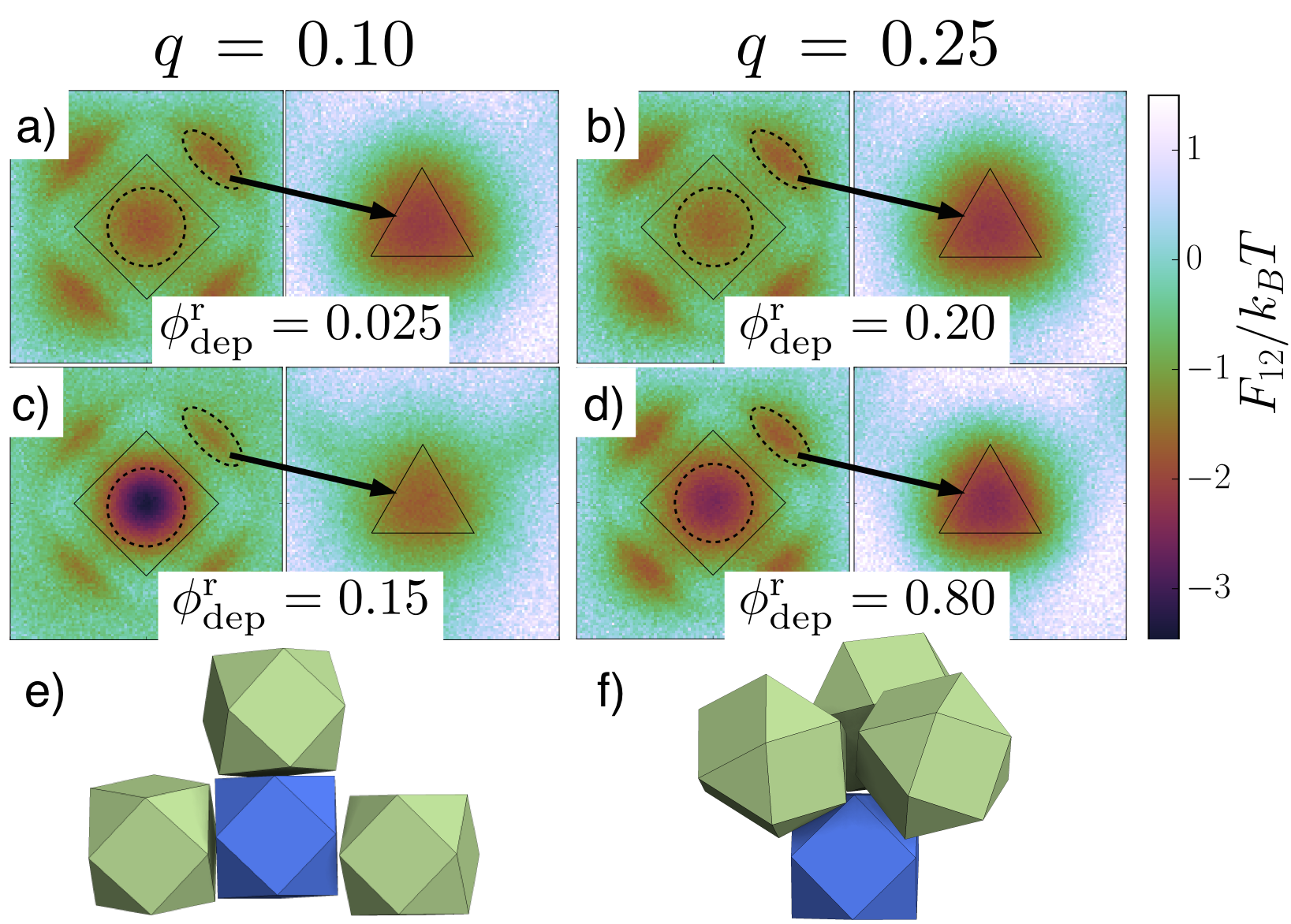}
\caption{ 2D slices of the PMFT through the first neighbor shells located next to both the square and triangular facets of the cuboctahedra for various depletant conditions. Calculations were performed for system sizes of $N=1000$ at $\phi_{\mathrm{col}}=0.55$. Each column uses depletants of different sizes, with \(q=0.10\) in the left column and \(q=0.25\) in the right column. For the PMFT across the square facet, it is possible to see the energy well along one of the square facets as well as four energy wells associated with different triangular facets. For the PMFT along the triangular facet, only the single energy well is visible. \textbf{(a \& b)} Under conditions that lead to formation of a rotator BCC crystal, the DEFs along the square facets are $ > -2 kT $, and the DEFs along the triangular facets are $ \approx -2 kT$. \textbf{(c)} For conditions that lead to formation of the SC crystal, the average DEFs are $-3.2 kT$ at the center of the square facets and $ -1.7 kT $ at the center of the triangular facets. \textbf{(d)} For conditions that lead to formation of the sBCC crystal, the average DEFs are $-2.4 kT$ at the center of the square facets and $ -2.3 kT $ at the center of the triangular facets.\textbf{(e \& f)} Representative configurations for a reference particle (blue) and a selection of its nearest neighbors (green). \textbf{(e)} The low coordination bonding along the large facets that is more likely to occur when \(q=0.1\). \textbf{(f)} The high coordination bonding that occurs when \(q=0.25\) and which leads to sBCC crystal nucleation.}
\label{pmft}
\end{figure}

The results presented so far speak to the system-wide behavior and the balance between colloid and depletant entropy. They do not, however, provide information on how depletion changes the behavior of the colloidal particles within their local environment. In order to understand these local environment changes, we have calculated the PMFT in the dense fluid state. The PMFT calculations shown in Figure \ref{pmft} depict the DEFs on both the six large, square facets and the eight smaller, triangular facets of the cuboctahedra, averaged out across ten unique simulation trajectories at a density \(\phi_{\mathrm{col}}=0.55\) before any noticeable crystals formed. The figures depict a two-dimensional slice through the first neighbor shell directly next to these facets. We ran these calculations for the two separate depletant-to-cuboctahedra size ratios, \(q=0.10\) and \(q=0.25.\) For both ratios, we considered depletant concentrations that will lead to formation of a rotator crystal and a non-rotator crystal so as to capture the differences of the three separate self-assembly regimes.

Figure \ref{pmft} illustrates how the emergent DEFs that lead to crystal assembly depend upon the depletant conditions. In Figures \ref{pmft}\textbf{(a \& b)}, the PMFT shows only moderate DEFs at both the square and triangular facets. At these low depletant densities, the driving force for facet alignment is weak. In Figure \ref{pmft}\textbf{c}, the high depletant concentration for $q=0.10$ leads to significantly stronger DEFs at the square facets and a reduction in DEFs at the triangular facets; thus there is a strong driving force for the square facets to align. In the conditions that will lead to the sBCC crystal depicted in Figure \ref{pmft}\textbf{d}, the DEFs on the square facets increase substantially while there is also a slight increase in the forces on the triangular facets. These conditions make alignment along all facets more favorable as compared to the depletant-free system.

From analyzing these PMFTs, we infer that depletion can change the effective particle shape when we consider hard particle assembly as being driven by directional entropic forces. As a general rule, increasing the size of a particle facet increases the strength of its directional entropic forces \cite{Damasceno2012b, vanAnders2014e, vanAnders2014d}. Since depletion increases the amount of crowding in the system, it is no surprise that the addition of depletants can increase DEFs and hence make particles have effectively larger facets. However, it is a remarkable result that depletants increase the DEFs on the large facets while decreasing the DEFs on the smaller facets. In this manner, depletants have the ability to make cuboctahedra effectively behave as cubes.

\section{Conclusion}

The ability to adjust bonding specificity via depletion size has ample opportunity to be used in controlling colloidal assemblies. van Anders et al.~\cite{vanAnders2014e} showed how increasing the facet size of truncated spheres will increase the specificity of DEFs. Here we achieve a similar effect in a different way, by adjusting depletant size, with smaller depletants increasing the specificity of DEFs. Thus, depletion can alter the effective shape of anisotropic particles due to the enhanced anisotropy of the DEFs with small depletants. We expect this knowledge to be useful for experiments because it is far easier to adjust assembly behavior by adding depletants or shrinking/swelling polymer coils compared to producing particles with a shape that can be varied \textit{in situ}. Overall, we view the additional design knobs of depletant size and concentration as simple, powerful tools to increase the assembly possibilities of anisotropic particles.

\section{Acknowledgements}
SCG acknowledges early discussions with G. van Anders and N.K. Ahmed on controlling effective particle shape via depletants. We thank A. Haji-Akbari for a helpful discussion regarding free energy. We additionally thank G. van Anders, J. Dshemuchadse, E. Harper, and X. Du for input and critical comments on the manuscript. This material is based upon work supported in part by the U.S. Army Research Office under Grant Award No. W911NF-10-1-0518 and also by a Simons Investigator award from the Simons Foundation to SCG.  This research was supported in part through computational resources and services supported by Advanced Research Computing at the University of Michigan, Ann Arbor, and used the Extreme Science and Engineering Discovery Environment \cite{TownsXSEDE} (XSEDE), which is supported by National Science Foundation grant number ACI-1053575; XSEDE award DMR 140129.




\bibliographystyle{rsc} 

\end{document}


\renewcommand*\rmdefault{bch}\normalfont\upshape
\rmfamily
\section*{}

\title{ \textit{Supplementary Information} \\Using Depletion to Control Colloidal Crystal Assemblies of Hard Cuboctahedra}
\author{Andrew S. Karas}
\affiliation{Department of Chemical Engineering, University of Michigan, 2800 Plymouth Rd. Ann Arbor, MI 48109, USA}
\author{Jens Glaser}
\affiliation{Department of Chemical Engineering, University of Michigan, 2800 Plymouth Rd. Ann Arbor, MI 48109, USA}
\author{Sharon C. Glotzer}
\email{sglotzer@umich.edu}
\affiliation{Department of Chemical Engineering, University of Michigan, 2800 Plymouth Rd. Ann Arbor, MI 48109, USA}
\affiliation{Department of Materials Science and Engineering, University of Michigan, 2300 Hayward St. Ann Arbor, MI 48109, USA}
\affiliation{Biointerfaces Institute, University of Michigan, 2800 Plymouth Rd. Ann Arbor, MI 48109, USA}
\maketitle
\nopagebreak

\section{Formation of simple cubic crystal with \(q=0.07\)}

Henzie et al. \cite{Henzie2012b} previously studied the sedimentation-driven assembly of cuboctahedra in systems with and without depletion. In their system, the cuboctahedra had an edge length of 145 nm and they employed a depleting polymer with a radius of gyration equal to 10 nm; this corresponds to \(q=\sigma_{\mathrm{dep}} / \sigma_{\mathrm{col}} = 0.069\). In experiments and simulations employing effective potentials to capture depletion effects, they observed cuboctahedra forming a face-centered tetragonal structure with I4/mmm symmetry. They reported this structure to form in Monte Carlo simulations when effective potentials represented penetrable hard sphere depletants at a packing fraction within the system of 0.03.

\begin{figure}[b]
\centering
\includegraphics[width=\textwidth]{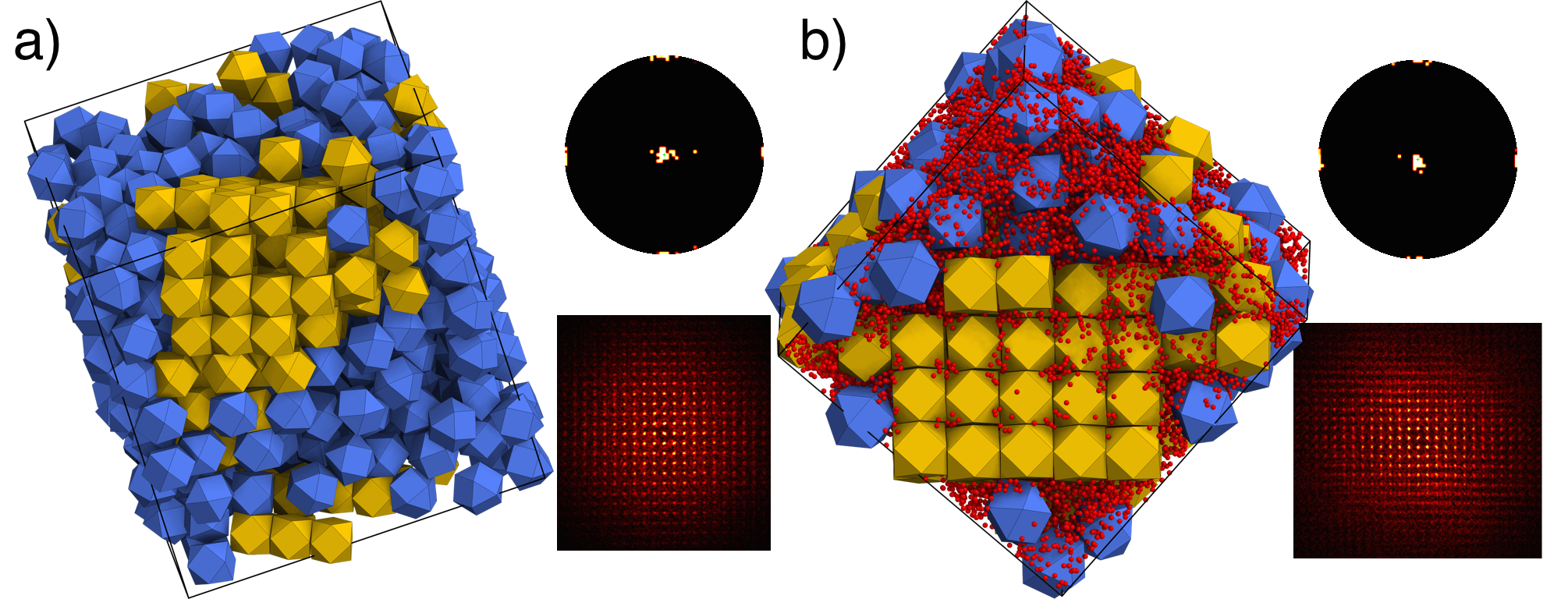}
\caption{The assembly of a simple cubic crystal using \textbf{(a)} the implicit depletion method and \textbf{(b)} explicit depletants. In both cases, the depletants have \(q=0.07\) and the concentration is set so that the depletant packing fraction in the system is 0.03. Particles in the solid are colored yellow, and identified using a Q4 spherical harmonic order parameter\cite{Steinhardt1983}. Additionally, we include a bond order diagram of a single crystal cluster and a diffraction pattern of the solid particles viewed along one of the four-fold axes.}
\label{assembly}
\end{figure}

We have tested our simulations against their results by using the implicit depletant method \cite{JensMethods} as well as by explicitly simulating penetrable hard sphere depletants, a significantly more laborious calculation. Figure \ref{assembly}(a,b) shows cuboctahedra forming a simple cubic crystal with \(q=0.07\) in simulations with implicit and explicit depletants. In both simulations, the colloid density is set to \(\phi_{\mathrm{col}} = 0.50\). The simulation with implicit depletants contains \(N_{\mathrm{col}}=512\) with \( \phi_{\mathrm{dep}}^{\mathrm{r}} = 0.08\), which corresponds to a system depletant packing fraction of 0.03. The simulation with explicit depletants contains \(N_{\mathrm{col}}=216\) and a packing fraction of depletants equal to 0.03. That we observe an SC crystal forming under both implicit and explicit treatments of depletants leads us to believe that gravity or some other force that our model does not include leads to the formation of the previously reported face-centered tetragonal structure.

\section{Gas - Liquid Coexistence}

\begin{figure}[t]
\centering
\includegraphics[width=0.6\textwidth]{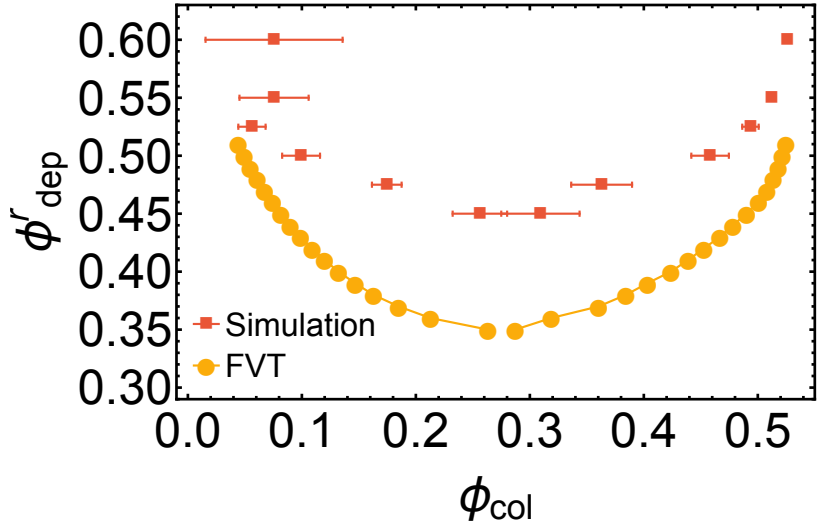}
\caption{Coexistence curves for gas-liquid phase separation observed in simulation and predicted from Free Volume Theory. For the simulations, we employ a Gibbs ensemble with a total $N=250$ cuboctahedra at an initial packing fraction $\phi_{\mathrm{col}} = 0.275.$ Simulation data was averaged across four independent runs.}
\label{gibbs}
\end{figure}

In order to further confirm the phase behavior predicted from Free Volume Theory, we ran simulations to observe gas-liquid coexistence. The standard procedure for doing this is to run simulations in the Gibbs ensemble \cite{FrenkelSmit}. In the Gibbs ensemble, we simulate two systems which are allowed to exchange volume and particles with one another, but the total volume and number of particles is held constant. The results from these simulations are summarized in Figure \ref{gibbs}. As we observed with fluid-crystal coexistence, theory tends to underestimate the critical depletant concentration needed for phase separation. Our implementation of the Gibbs ensemble is described in Ref. \cite{JensMethods}.

\section{Free Energy Calculations}

In a semigrand ensemble, such as the one we consider, the full expression for the free energy is: \cite{Lekkerkerker2011b}

\begin{equation}
F (N_{\mathrm{col}}, \mu_{\mathrm{dep}}, V, T) = F(N_{\mathrm{col}}, \mu_{\mathrm{dep}} \rightarrow - \infty, V, T) - \int_{- \infty}^{\mu_{\mathrm{dep}}} N_{\mathrm{dep}} d\mu_{\mathrm{dep}}
\end{equation}

Here, $N_{\mathrm{dep}}$ is the number of depletants in the system. According to the Widom insertion theorem \cite{WidomInsertion}, the chemical potential of depletants in the system is $\mu_{\mathrm{dep}} = \mathrm{const} + kT \mathrm{ln} (N_{\mathrm{dep}} / \langle V_{\mathrm{free}} \rangle ) $ and, by definition, the chemical potential in the external reservoir is $\mu_{\mathrm{dep}} = \mathrm{const} + kT \mathrm{ln} (n_{\mathrm{dep}}^{\mathrm{r}}) $. Equating these terms leads to the result $N_{\mathrm{dep}} = n_{\mathrm{dep}}^{\mathrm{r}} \langle V_{\mathrm{free}} \rangle$. Substituting these expressions into the free energy relationship, and making the critical assumption that $\langle V_{\mathrm{free}} \rangle$ can be treated as a constant as the depletion concentration varies leads to:

\begin{equation}
F (N_{\mathrm{col}}, \mu_{\mathrm{dep}}, V, T) \approx A(N_{\mathrm{col}}, V, T) - n_{\mathrm{dep}}^{\mathrm{r}} \, \langle V_{\mathrm{free}} \rangle \, kT
\end{equation}

$\langle V_{\mathrm{free}} \rangle$ can be replaced by definition with $\alpha V$. In doing our analysis with a common tangent construction, we work with the free energy density $ f (\phi_{\mathrm{col}}, \mu_{\mathrm{dep}}) = F (N_{\mathrm{col}}, \mu_{\mathrm{dep}}, V) v_{0} / V $ where $ v_{0} $ is the volume of a single colloid. The purpose of the common tangent construction is to find densities at which two phases will coexist, meaning that they have an equal pressures and chemical potentials. The pressure is defined as:

\begin{equation}
P = - \left[\frac{\partial F}{\partial V}\right] = - \frac{1}{v_{0}} \left[ \frac{\partial (V f)}{\partial V}\right]  =  - \frac{1}{v_{0}} \left[ f - \phi_{\mathrm{col}} \frac{\partial f}{\partial \phi_{\mathrm{col}}}\right] 
\end{equation}
The chemical potential is:

\begin{equation}
\mu = \frac{\partial F}{\partial N} = \frac{1}{v_{0}} \left[ \frac{\partial (V f)}{\partial N} \right] = \frac{\partial f}{\partial \phi_{\mathrm{col}}}
\end{equation}
For more on common tangent construction use with FVT, we refer the interested reader to reference \cite{Dijkstra1999}.

As mentioned earlier, we rely on numerical calculations of the free energy. For the SC and sBCC crystals, we computed the Helmholtz free energy through Frenkel-Ladd integration \cite{Frenkel1984a} of the bulk solid using the method described by Haji-Akbari et al.~\cite{Haji-Akbari2011h}. This implementation for anisotropic particles considers both the translational and rotational degrees of freedom by tethering particles to springs about their average positions and orientations in the lattice. We performed the calculations on bulk SC and sBCC crystals for a variety of different crystal densities, \(\phi_{\mathrm{col}}\), up to the densest packings (\(\sim 0.833\) for SC and \(\sim 0.917\) for sBCC).

We rely on data from the cuboctahedra equation of state when calculating free energies of the other phases. We first ran NPT simulations of systems with 4096 particles. From this, we computed the compressibility factor $ Z = PV / NkT $. We then used Widom particle insertion \cite{WidomInsertion} to determine the chemical potential of the system at a reference density of $ \phi_{\mathrm{col, 0}} = 0.20 $. We calculate the chemical potential at the reference density as: 
\begin{equation}
\mu_{\mathrm{0}}(\phi_{\mathrm{col, 0}}) = \mathrm{ln} (\phi_{\mathrm{col, 0}}) - \mathrm{ln}(\mathrm{B}_{i}) -  \mathrm{ln}(2\pi^2)
\end{equation}
where $\mathrm{B}_i$ is the probability of randomly inserting a particle without generating an overlap, and the last term is necessary to make the normalization  of the fluid free energy consistent with the free energy computed in our Frenkel-Ladd calculations. Because we express rotations using normalized quaternions $\mathbf{q}^2 = 1$, the factor of $2 \pi^2$ measures the surface
of the 3-sphere.

This value for the chemical potential at a reference density allows us to determine the absolute free energy by integrating over the compressibility factor:

\begin{equation}
f (\phi_{\mathrm{col}}) = \phi_{\mathrm{col}} \bigg[ \, \mu_{0}(\phi_{\mathrm{col, 0}}) \, - \, Z_{0}(\phi_{\mathrm{col, 0}}) +  \int_{\phi_{\mathrm{col, 0}}}^{\phi_{\mathrm{col}}} \frac{Z(\phi_{\mathrm{col}})}{\phi_{\mathrm{col}}} d \phi_{\mathrm{col}} \, \bigg]
\end{equation}

We similarly integrate over the equation of state to describe the rotator BCC phase and part of the sBCC phase. When integrating the equation of state in these cases, we use an integration constant in order to match the pressure and chemical potential for the unperturbed fluid-rotator BCC coexistence and the rotator BCC-sBCC coexistence from reference \cite{Gantapara2013}.

\begin{figure}
\centering
\includegraphics[width=\textwidth]{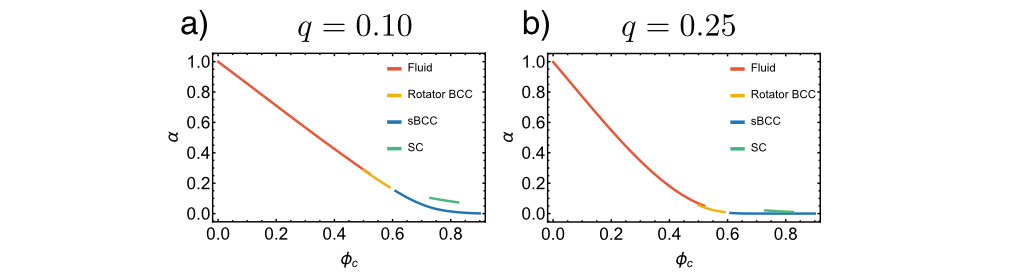}
\caption{Plots of $\alpha = \langle V_{\mathrm{free}} \rangle / V$ for \textbf{(a)} $q=0.10$ and \textbf{(b)} $q=0.25$. As $q$ increases, the $\alpha$ will go to a value of zero for both the SC and sBCC structures.}
\label{alpha}
\end{figure}

To calculate the depletant contribution to free energy, we used Monte Carlo integration to determine the free volume as a function of depletant size across different colloid densities and phases and in both the SC and sBCC crystals. This allowed us to have an expression for $\alpha = \langle V_{\mathrm{free}} \rangle / V$ across the full range of colloid densities (pictured in Figure \ref{alpha}).

\begin{figure}
\centering
\includegraphics[width=\textwidth]{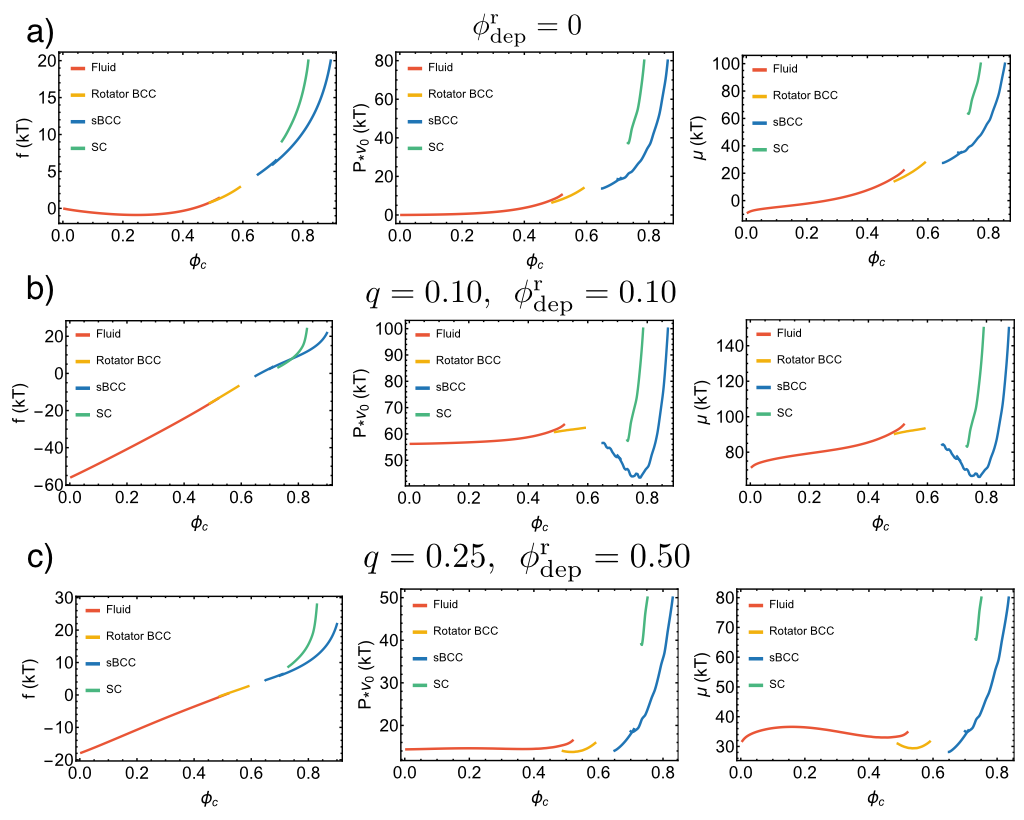}
\caption{Plots showing the free energy density, pressure, and chemical potential in the fluid, rotator BCC, sBCC, and SC phases at different depletant conditions. \textbf{(a)} With no depletants, the SC phase is never stable and the sBCC phase is only stable at high $\phi_{\mathrm{col}}$. \textbf{(b)} At $q=0.1 \, \, \& \, \, \phi_{\mathrm{dep}}^{\mathrm{r}}=0.10$, the fluid-SC coexistence becomes stable.  \textbf{(c)} At $q=0.25 \, \, \& \, \, \phi_{\mathrm{dep}}^{\mathrm{r}}=0.50$, the system will phase separate into a fluid and sBCC crystal.}
\label{free_energy}
\end{figure}

We include calculations of the free energy density $f$, the pressure, and chemical potential for three depletant conditions in Figure \ref{free_energy}.

\bibliographystyle{rsc}

\providecommand*{\mcitethebibliography}{\thebibliography}
\csname @ifundefined\endcsname{endmcitethebibliography}
{\let\endmcitethebibliography\endthebibliography}{}